# Ballistic transport of graphene *pnp* junctions with embedded local gates

**Seung-Geol Nam[1], Dong-Keun Ki[1] ‡, Jong Wan Park[2], Youngwook Kim[1], Jun Sung Kim[1], and Hu-Jong Lee[1] §**

[1] Department of Physics, Pohang University of Science and Technology, Pohang 790-784, Republic of Korea

[2] Technology Development Team, National NanoFab Center, Daejon 305-806, Republic of Korea

E-mail: `hjlee@postech.ac.kr`

**Abstract.**
   We fabricated graphene *pnp* devices, by embedding pre-defined local gates in an oxidized surface layer of a silicon substrate. With neither dielectric-material deposition nor electron-beam irradiation on the graphene, we obtained high-quality graphene *pnp* devices without degradation of the carrier mobility even in the local-gate region. The corresponding increased mean free path leads to the observation of ballistic and phase-coherent transport across a 130-nm-wide local gate, which is about an order of magnitude wider than reported previously. Furthermore, in our scheme, we demonstrated independent control of the carrier density in the local-gate region, with a conductance map very distinctive from top-gated devices. It was caused as the electric field arising from the global back gate is strongly screened by the embedded local gate. Our scheme allows the realization of ideal multipolar graphene junctions with ballistic carrier transport.

‡ Present address: DPMC and GAP, University of Geneva, CH1211 Geneva, Switzerland
§ Author to whom any correspondence should be addressed.

# 1. Introduction

Graphene, a single atomic layer of graphite, has attracted considerable interest owing to its unique linear band structure as well as its promising device applications [1]. Due to its gapless energy spectrum, low carrier density, and atomic thickness, the carrier type and density can be conveniently tuned by using electrostatic gates, which allows the *in-situ* realization of a *pn-* or *pnp*-type multipolar configuration on a graphene layer. A graphene *pnp* junction offers a unique platform for investigating many novel phenomena such as Klein tunneling [2], particle collimation [3], and Veselago lensing [4].

To realize electronic devices based on these unusual transport properties in graphene heterostructures, fabrication of a high-quality graphene multipolar junction is essential. In most of the studies to date, graphene *pnp* junctions are fabricated by placing a top local gate separated from the underlying graphene layer by a metal-oxide or an organic dielectric layer [5, 6]. However, in this scheme, charged impurities left in such dielectrics provide additional doping and scattering sites, thereby limiting the carrier mobility in the graphene [7]. Moreover, since a top-gated graphene sheet is covered with dielectrics an annealing process [8, 9], which is often adopted to enhance the carrier mobility by removing the organic residue in the graphene layer, is not applicable. To overcome such difficulties, air-bridged top gates have been attempted [10, 11]. However, due to the large vacuum gap between the suspended gate and the underlying graphene sheet, an excessively high voltage must be applied to induce a sufficient carrier density, which is not practical for device applications. Furthermore, at high voltage, the electrostatic force tends to collapse the suspended gate, thus restricting the range of possible lateral sizes and thicknesses of the gates to ensure the mechanical strength of a device. Finally, placing a top gate requires electron-beam irradiation of the surface of the graphene, either to cross-link the organic resist or to define the local gates, which tends to degrade the graphene layer [12, 13, 14].

In this paper, we introduce a new method of fabricating graphene *pnp* junctions. By embedding pre-patterned local gates in a substrate, we are able to obtain high-quality *pnp* junctions with neither dielectric-layer deposition nor electron-beam exposure of the graphene sheet. Similar schemes have been adopted to realize high-quality local-gated carbon nanotube devices [15, 16]. We achieved ballistic and phase-coherent carrier transport in a graphene *pnp* device with a 130-nm-wide local gate, which is almost an order magnitude wider than reported previously [17]. Quantitative analysis using the Landauer formula explains the observed conductance oscillation well and shows that the mobility is not degraded in the local-gate region (LGR). In a high magnetic field, another device with a 1-$\mu$m-wide local gate exhibited the $2e^2/h$ quantum-Hall (QH) plateau, indicating no backscattering in the LGR. The conductance across our *pnp* junctions shows a gate-voltage dependence that is very distinctive from that of top-gated junctions, indicating a strong screening of the electric field by the embedded local gate.

## 2. Sample preparation and measurements

Figure 1(a) illustrates the device fabrication process where poly-silicon gates were embedded in the silicon-oxide layer. The process begins with the deposition of an 80-nm-thick poly-silicon layer on top of a highly doped silicon wafer covered with a 200-nm-thick silicon-oxide layer. The poly-silicon layer was made conductive by implanting 40 keV phosphorus ions with a dose of $5.0 \times 10^{15}$ cm$^{-3}$. This highly doped poly-silicon was then etched into a pre-designed pattern using the negative electron-beam resist (NER) as an etching mask. After depositing an additional silicon-oxide layer on the entire wafer, chemical-mechanical planarization (CMP) was used to obtain a flat surface. At this stage, the silicon-oxide layer was etched until the poly-silicon surface appeared. Finally, an approximately 10-nm-thick silicon-oxide layer was deposited. Figure 1(b) shows a cross-sectional transmission electron microscopy (TEM) image of the final product where conducting poly-silicon wires, to be used for local gates, are buried $\sim 10$ nm below the surface of the silicon-oxide layer.

Figure 1(c) shows optical images of graphene sheets placed on embedded local gates with widths of 1 $\mu$m (upper panel) and 130 nm (lower panel). A graphene sheet was mechanically transferred onto a local gate [18], enabling the placement of a graphene sheet within a few $\mu$m of the intended position. The transfer process proceeded by first exfoliating a single-layer graphene onto a polymer bilayer consisting of poly (4-styrenesulfonic acid) at the bottom and polymethyl methacrylate (PMMA) at the top. After a suitable piece of graphene was selected using an optical microscope, the polymer stack was floated on the surface of a deionized water bath. Since poly (4-styrenesulfonic acid) is water-soluble, only the PMMA membrane and the graphene layer remained on top of the water. The graphene on PMMA membrane was then transferred to a target substrate with embedded local gates [18]. The graphene sheet was positioned within a few micrometers from the intended position using a micromanipulator mounted under an optical microscope. Note that the graphene surface was kept intact by avoiding chemical contact with any solvents during the transfer process. After the transfer, the substrate was heated to 110 °C for 10 minutes to enhance the substrate-graphene adhesion. The PMMA was then dissolved away in acetone, and electrodes were attached to the graphene by electron-beam patterning and the subsequent evaporation of a Ti/Au (5/35 nm in thickness) bilayer. An optical image and a schematic diagram of a typical device are shown in figures 2(a) and 2(b), respectively.

We made measurements in a dilution fridge in a temperature range of 60 mK − 4.2 K with magnetic fields up to 10 T. Conventional lock-in techniques were used with a current bias of 10-500 nA at 13.3 Hz. Results from two representative devices are presented here; the data in figures 2 and 3 were taken with the device shown in figure 2(a), which had a 1-$\mu$m-wide local gate and a carrier mobility of $\sim 5,000$ cm$^2$/Vs in graphene, and the data in figure 4 were taken with a device having a 130-nm-wide local gate and a carrier mobility of $\sim 8,000$ cm$^2$/Vs (corresponding to the mean free path of $\sim 165$ nm), as determined at the carrier density of $n \sim 3.2 \times 10^{12}$ cm$^{-2}$. The widths of the local gates were determined using a scanning electron microscope.

## 3. Results and discussion

To specify the number of atomic layers of adopted graphene, the QH resistance $R_{xy}$ in a global-gate region (GGR) was measured at $B = 10$ T. As shown in figure 3(a), QH plateaus with values of 1/2, 1/6, and 1/10 in units of $h/e^2$ were observed, which are characteristic of single-layer graphene [1]. The longitudinal resistance $R_{xx}$ showed successive peaks with intervals of $\sim 12.6$ V as $V_{BG}$ varied. Since the degeneracy of the Landau levels at $B = 10$ T is $\sim 10^{12}$ cm$^{-2}$, we estimated the corresponding back-gate coupling efficiency, $\alpha_{BG}=n_1/V_{BG} \approx 7.9 \times 10^{10}$ cm$^{-2}$/V, where $n_1$ is the carrier density in the GGR. Unlike top-gated devices, electrodes for the measurement of QH states in the LGR can be conveniently placed on the LGR. In figure 3(b), QH resistance in the LGR is shown as a function of $V_{LG}$, where half-integer QH plateaus are clearly seen. The local-gate coupling efficiency extracted from the interval between the center of plateaus, $\sim 0.53$ V, gives $\alpha_{LG} = n_2/V_{LG} \approx 1.9 \times 10^{12}$ cm$^{-2}$/V with $n_2$ being the carrier density in the LGR. $\alpha_{BG}$ and $\alpha_{LG}$ agree reasonably well with the estimate from the simple parallel-plate-capacitor model, $e\alpha_{LG(BG)} = \varepsilon_0\varepsilon_{SiO_2}/h_{1(2)}$, together with the dielectric constant of silicon oxide $\varepsilon_{SiO_2} = 3.9$ and a spacing between the local (back) gate and the graphene of $h_{1(2)} =$11 (270) nm.

In figure 2(c), the four-terminal resistance across the *pnp* junction is plotted as a function of the back-gate voltage ($V_{BG}$) and the local-gate voltage ($V_{LG}$). The most prominent feature is the upright cross-like pattern that divides the conductance map into four quadrants. It consists of vertical and horizontal bands, which correspond to the charge neutrality points (CNP) in the LGR and the GGR, respectively. $R(V_{LG}, V_{BG})$, shown in figure 2(c), is in clear contrast to that observed in top-gated devices, where a skewed cross-like pattern was observed [6, 19, 20]. The difference is more clearly seen in a one-dimensional slice plot at a fixed $V_{LG}$ or $V_{BG}$. In top-gated devices, where the local gate is located above the graphene sheet, the carrier density in the LGR is affected by both $V_{LG}$ and $V_{BG}$. Thus, tuning the charge-neutral condition in the LGR depends on both $V_{LG}$ and $V_{BG}$. This leads to a doubly peaked structure in $R(V_{BG})$ [19, 20], and a varying peak position in $R(V_{LG})$ [6].

In our device, however, neither double-peak structures in $R(V_{BG})$ nor the change in the peak position of $R(V_{LG})$ were observed [see figures 2(d) and 2(e)]. This indicates that the carrier density in the LGR is not affected by $V_{BG}$, but depends on $V_{LG}$ alone. We attribute this behavior to the poly-silicon conducting wire (local gate) located between the graphene and the back gate, which almost completely screens the electric field induced by the back gate. To confirm this interpretation, we calculated the carrier density profile near the LGR using commercial finite-element simulation software (Comsol Multiphysics). In the calculation, we assumed that the graphene was perfectly conducting and used the potential of the local and back gates as boundary conditions. We chose 3.9 for the dielectric constant of silicon oxide and set the spacing between the graphene layer and the local gate (back gate) at 11 nm (270 nm).

Figure 2(f) shows the profile of the carrier density in graphene under a 1-$\mu$m-wide local gate, calculated at $V_{LG} = 4$ V for different values of $V_{BG}$. Since the spacing between the local gate and the graphene layer in our device is only 11 nm, the electric potential between graphene and the local gate is strongly influenced by the local gate voltage itself but not by the voltage of the back gate that is located 270 nm below the graphene layer. The back-gate effect is almost completely screened out by the conducting local gate. On the other hand, as the electric field decreases rapidly moving away from the narrow local gate, the electric potential is little influenced by the local gate voltage and is determined solely by the global back gate voltage. This explains why the carrier density in the LGR ($-0.5$ $\mu$m$< x <0.5$ $\mu$m) is constant for all values of $V_{BG}$ in figure 2(f), consistent with our expectations and the experimental results shown in figures 2(c), 2(d), and 2(e).

The asymmetry in the resistance profiles in figures 2(d) and 2(e) indicates the formation of a *pnp* multipolar configuration in graphene, because a *pn* (bipolar) interface yields a higher resistance than an *nn* (unipolar) interface [6]. The formation of a *pn* interface can also be confirmed by transport measurements in the QH regime. In a high magnetic field, the QH effect in the GGR and LGR develops independently with the separate respective filling factors of $\nu_1$ and $\nu_2$. Mixing of the edge states at the LGR-GGR boundary leads to quantized two-terminal conductance at a fraction of $e^2/h$ [5, 19, 21]. In a *pnp* junction, the two-terminal conductance $G$ depends on the relative magnitude and sign of $\nu_1$ and $\nu_2$ [19].

(a) edge-state transmission regime ($|\nu_1| > |\nu_2|$ and $\nu_1\nu_2 > 0$),
$$G = e^2/h|\nu_2| \tag{1}$$
(b) partial-equilibrium regime ($|\nu_1| < |\nu_2|$ and $\nu_1\nu_2 > 0$),
$$G = \frac{e^2}{h}\frac{|\nu_1||\nu_2|}{2|\nu_2| - |\nu_1|} \tag{2}$$
(c) full-equilibration regime ($\nu_1\nu_2 < 0$),
$$G = \frac{e^2}{h}\frac{|\nu_1||\nu_2|}{2|\nu_2| + |\nu_1|} \tag{3}$$

Figure 3(c) presents a color map of the conductance plateaus calculated from Eqs. (1)-(3) for different values of $\nu_1$ and $\nu_2$.

Figure 3(d) displays the measured two-terminal conductance of our device as a function of $V_{BG}$ and $V_{LG}$ for $B = 10$ T. The conductance map consists of adjoined rectangles, which correspond to different combinations of $\nu_1$ and $\nu_2$. Again, due to the strong screening effect discussed above, the shape of the conductance map in figure 3(d) differs in its principal feature from top-gated devices, where adjoined parallelograms were observed [5, 19]. The feature in figure 3(d) is in qualitative agreement with that calculated in figure 3(c). To present a quantitative comparison, we plotted the conductance as a function of $V_{LG}$ for $\nu_1 = -2$ (upper panel) and $-6$ (lower panel) in figure 3(e). The values of the conductance at the plateaus are denoted in figure 3(c) in units of $e^2/h$. In both graphs, the measured conductances are in excellent agreement with the calculations, which are denoted by broken lines. In particular, for $\nu_1 = \nu_2 = -2$, a quantized conductance plateau $2e^2/h$ is clearly visible, which was previously observed only in devices with suspended top gates [11]. The absence of this plateau in ordinary top-gated devices, *i.e.*, those with a dielectric layer between the graphene and the top gate, is presumably due to backscattering in the LGR and/or the *pn* interfaces [19]. Thus, the presence of the $2e^2/h$ plateau in our device indicates improved junction quality and no backscattering.

We next focus on the behavior of the device in the GGR with a 130-nm-wide local gate and a mean free path of $\sim 165$ nm. To observe ballistic as well as phase-coherent electron transport in a graphene *pnp* junction, the spacing between two adjacent *pn* interfaces should be shorter than the mean free path. This was first demonstrated by Young and Kim in a device with an extremely narrow local gate ($\leq 20$ nm) [17]. They observed Fabry-Perot interference of the carriers between the two *pn* interfaces of a *pnp* junction. More importantly, they also observed a $\pi$ phase shift in the Fabry-Perot oscillations in a finite magnetic field, which signifies Klein tunneling [2, 17, 22]. However, similar measurements conducted by other groups did not show the $\pi$ phase shift in the Fabry-Perot oscillations [10, 23], presumably because the ballistic transport condition was not satisfied in the LGR [24].

Figure 4(a) shows the local gate dependence of the conductance across the *pnp* junction at $V_{BG} = $ -40 V. When a *pnp* junction is established ($V_{LG} > V_{CNP} \sim -0.03$ V), the conductance oscillates as a function of $V_{LG}$. Around $V_{LG} = 1$ V, the spacing between successive peaks becomes $\sim 0.3$ V, corresponding to a carrier density of $\Delta n_2 \approx 5.7 \times 10^{11}$ cm$^{-2}$. This value agrees well with a naive estimate for the resonant density in a Fabry-Perot cavity [17], $\Delta n \approx 4\sqrt{\pi n_2}/L_c \sim 6.0 \times 10^{11}$ cm$^{-2}$, where $L_c$ is the spacing between the two *pn* interfaces. As shown in figure 4(b), the oscillations become clearer by plotting the numerical derivative of the conductance, $dG/dV_{LG}$. This periodic feature was observed only in the presence of a *pnp*-type multipolar configuration. This oscillation feature depends primarily on $V_{LG}$, because the width of a *pnp* junction and the electric field at *pn* interfaces are mostly determined by $V_{LG}$ [see the carrier density profile in figure 4(e)].

In a magnetic field, the observed conductance oscillations shows an abrupt phase shift for $B \sim 250$–500 mT [see figure 4(c), 4(d), and 4(f)]. This phase shift is caused by electrons normally incident on the *pn* junctions ($k_y = 0$) as they gain an additional Berry's phase $\pi$ in an external magnetic field by enclosing the origin in momentum space [17, 22]. In our devices, the phase shift is expected to occur for $B^* = 2\hbar k_{y0}/eL_c = \sqrt{4\ln(3/2)\hbar eE/\pi(eL_c)^2 v_F} \sim 400$ mT, which is in good agreement with observation. Here, $k_{y0}$ is the transverse momentum of an electron, which gives its largest contribution to the Fabry-Perot resonance in zero magnetic field; $E$ is the electric field at the *pn* interfaces; and $v_F$ is the Fermi velocity. For a more detailed analysis, we extract the oscillating component of conductance $G_{osc}$ by subtracting the symmetric and non-oscillating background conductance following the scheme adopted in previous studies [17]. We first obtained the odd component of the resistance by antisymmetrizing the resistance, $R_{odd}(|n_2|) = R(n_2) - R(-n_2)$. We then subtracted the background conductance, $\overline{G_{odd}}$ from the odd conductance component, yielding $G_{osc}=G_{odd} - \overline{G_{odd}}$ as a function of $B$ and $V_{LG}$; see figure 4(d).

For a quantitative analysis, we calculated $G_{osc}$ by adopting the Landauer formula [17, 22],

$$G_{osc} = 8e^2/h \sum |T_R|^2 |T_L|^2 |R_R||R_L| \cos(\theta_{WKB} + \theta_{rf}) e^{-2L/\ell_{mfp}}, \qquad (4)$$

where $T_{R(L)}$ and $R_{R(L)}$ are the transmission and reflection amplitudes of the carriers at the right (left) side of the *pn* boundaries, $\theta_{WKB}$ is the phase accumulated across a *pnp* junction, $\theta_{rf}$ is the additional phase acquired upon back-reflection at a *pn* interface, and $\ell_{mfp}$ is the mean free path in the LGR, a fitting parameter that controls the amplitude of the oscillation. $T_{R(L)}$ and $R_{R(L)}$ are determined by the electric field at the *pn* interfaces [17, 22], which is given by $eE = 2.4\hbar v_F (n')^{2/3}$ with nonlinear screening at the CNP into account [25]. Here, $n'$ is the gradient of the carrier density at the junction boundary. To estimate $\theta_{WKB}$, $\theta_{rf}$, and $E$, we simulated the carrier density profile around the LGR using the device parameters of $W = 130$ nm, $h_1 = 11$ nm, and $h_2 = 270$ nm [see figure 4(e)]. For the mean free path $\ell_{mfp}$ in the LGR, we adopted the GGR value, which was extracted from the conductance measurement $\ell_{mfp} = (h/2e^2)\sigma/\sqrt{n_1\pi}$.

As shown in figure 4(f), with $\ell_{mfp} = 165$ nm, the calculated $G_{osc}$ is in excellent agreement with the data in both the zero and in the different finite magnetic fields. This indicates that the mean free path is almost identical in both the LGR and the GGR. This result contrasts with those of previous studies with top-gated devices, where the LGR mean free path turned out to be less than two thirds of that in the GGR [17]. The reduced mean free path in the LGR for those top-gated devices is attributed to the deposited dielectrics and the e-beam irradiation during the top-gate fabrication, which tends to degrade the LGR graphene layer [17, 26]. In our embedded local-gate devices, however, no variation occurred in the mean free path in the LGR and the GGR. The long mean free path in the LGR of our devices allowed the observation of ballistic phase-coherent transport, even with a 130-nm-wide local gate. This is an order of magnitude wider than those used in previous studies, where a top-gate width was less than 20 nm. This clearly demonstrates the merit of using embedded local gates and confirms the resulting high quality of our *pnp* junction.

## 4. Conclusion

In conclusion, by embedding pre-defined local gates on the surface of a substrate, we have developed a new scheme for fabricating graphene *pnp* devices. With neither dielectric-material deposition nor electron-beam exposure of LGR, we obtained high-quality *pn* junctions without degradation of the LGR mobility. This scheme allowed us to observe ballistic phase-coherent transport in a device with a local gate that was much wider than reported in a recent study [17]. The $2e^2/h$ QH plateau develops in high magnetic fields, indicating the formation of high-quality *pnp* junctions without backscattering.

With local gates imbedded inside the substrates, our device structure can be further improved by utilizing the freedom of adopting diverse methods of enhancing the mobility in graphene, such as Ar/$H_2$ annealing [9] or transference on boron-nitride substrates [18]. It enables the study of graphene *pnp* junctions in a fully ballistic regime, where one may be able to observe many novel phenomena such as the Veselago lensing effect [4]. Our scheme will also facilitate scanning-tunneling-spectroscopic studies [27] on graphene *pn* boundaries without dielectric coverage to probe the details of scattering and/or tunneling processes as well as the existence of snake states at finite magnetic fields [28, 29].

## Acknowledgments

SGN acknowledges the assistance of D. Jeong, G.-H. Lee, and S. Jo in measurements. This work was supported by the National Research Foundation (NRF) of Korea through Acceleration Research Grant R17-2008-007-01001-0 (for HJL) and through the Basic Science Research Program Grant 2009-0076700 (for JSK).

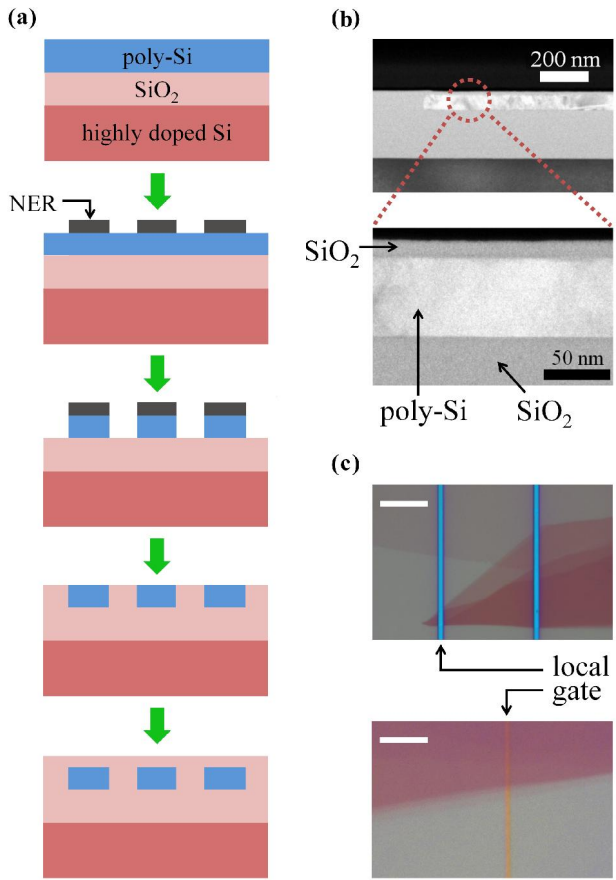

**Figure 1.** (Color online) (a) Fabrication processes for embedded local gates: deposition of a poly-silicon layer, depositing and patterning an etching-mask layer of negative electron-beam resist (NER), etching the poly-silicon layer, second deposition of silicon oxide, chemical-mechanical planarization, and the additional deposition of silicon oxide. (b) Cross-sectional view of a device with an embedded local gate. Images were taken with a transmission electron microscope. (c) Optical image of graphene positioned on top of the embedded local gates. The widths of the local gates are 1 $\mu$m (upper) and 130 nm (lower). Size of the scale bar is 10 (4) $\mu$m for the upper (lower) panel.

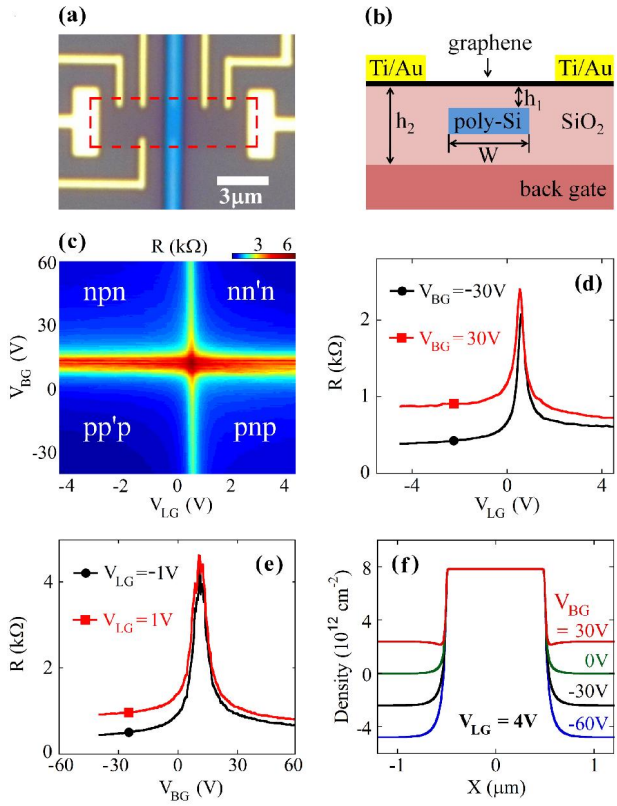

**Figure 2.** (Color online) (a) Optical image of a typical device with a 1-$\mu$m-wide local gate and metallic electrodes. The dashed line denotes the boundary of the graphene layer. (b) Schematic cross-sectional configuration of a typical device. (c) Resistance across the *pnp* junction ($R$) as a function of $V_{LG}$ and $V_{BG}$. (d) $V_{LG}$ dependence of the resistance at $V_{BG} = 30$ and -30 V. (e) $V_{BG}$ dependence of the resistance at $V_{LG} = 1$ and -1 V. (f) Calculated density profile around the local-gate region (LGR) at $V_{LG} = 4$ V for various $V_{BG}$.

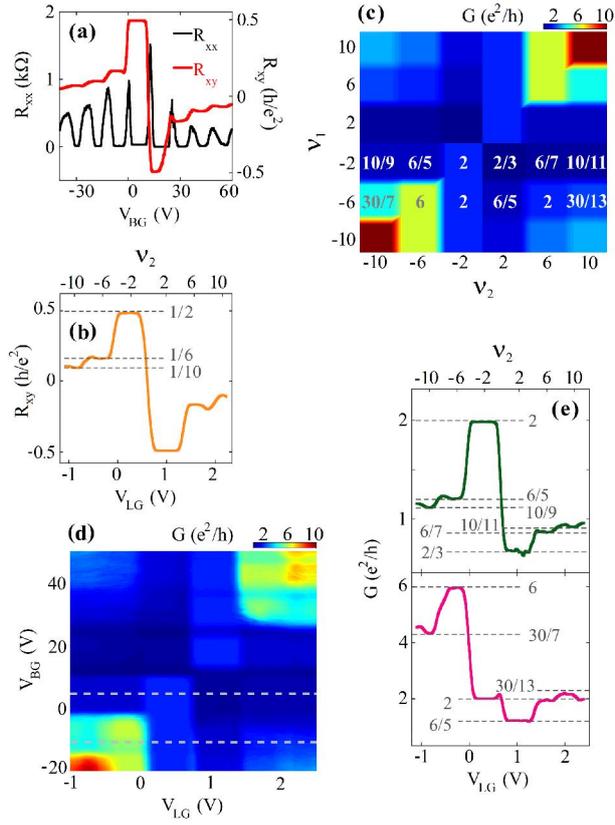

**Figure 3.** (Color online) (a) Longitudinal resistance ($R_{xx}$) and QH resistance ($R_{xy}$) in the global-gate region (GGR) in magnetic field $B = 10$ T. (b) QH resistance in the local-gate region in $B = 10$ T. (c) Calculated two-terminal conductance of the *pnp* junction ($G$) as a function of filling factors in the GGR ($\nu_1$) and the LGR ($\nu_2$). (d) Two-terminal conductance as a function of $V_{LG}$ and $V_{BG}$. (e) Local-gate dependence of the two-terminal conductance at $V_{BG}$ =7.2 V (upper) and -11 V (lower).

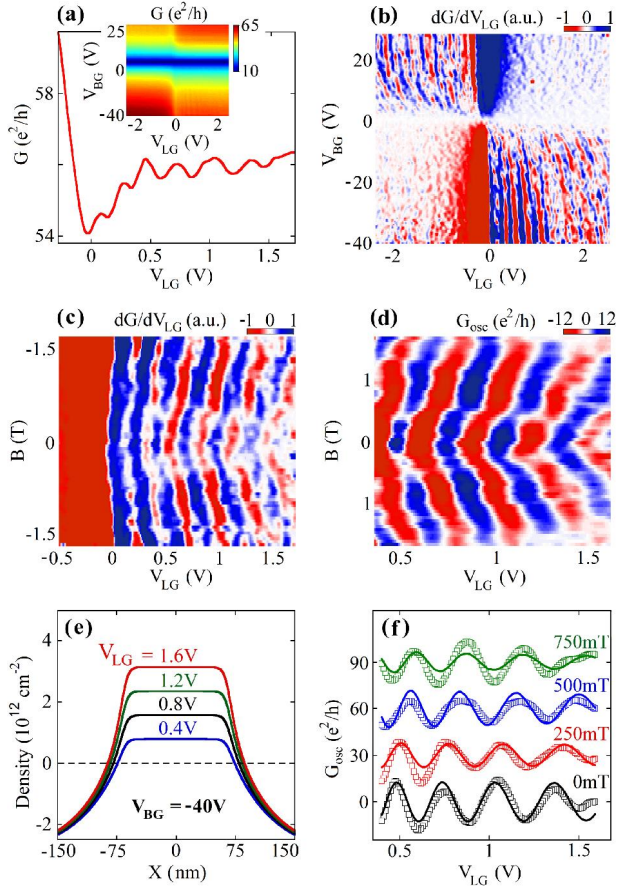

**Figure 4.** (Color online) (a) $V_{LG}$ dependence of the conductance ($G$) at $V_{BG}$ = -40 V. The inset shows the conductance as a function of $V_{LG}$ and $V_{BG}$. (b) Numerical derivative of the conductance, $dG/dV_{LG}$, as a function of $V_{LG}$ and $V_{BG}$. (c) $dG/dV_{LG}$ as a function of $V_{LG}$ and magnetic field $B$. (d) Oscillating component of the conductance $G_{osc}$ as a function of $V_{LG}$ and $B$. (e) Calculated carrier density around the LGR at $V_{BG} = -$ 40 V for various $V_{LG}$. (f) Measured (open rectangle) and calculated (solid line) $G_{osc}$ for various magnetic fields. Curves are offset for clarity.